\documentstyle[prb,aps,epsfig]{revtex}
   
   \def\lmax{{l_{\rm max}}}
      
   \def\rb{{\bf r}}   \def\vb{{\bf v}}   \def\Gb{{\bf G}}
\def\pb{{\bf p}}         \def\Qb{{\bf Q}}
\def\ee{{\rm e}}   
\def\ii{{\rm i}}       
    \def\Eb{{\bf E}}   
   \def\Ab{{\bf A}}   \def\jb{{\bf j}}   
      \def\eb{{\hat{\bf e}}}
   \def\Kb{{\bf K}}      \def\xh{{\hat{\bf x}}}
\begin{document}
\draft

\twocolumn[\hsize\textwidth\columnwidth\hsize\csname
@twocolumnfalse\endcsname

\title{Electron energy loss and induced photon emission in photonic crystals}

\author{F. J. Garc\'{\i}a de Abajo$^{1,2}$\cite{email}
        and L. A. Blanco$^{2}$}
\address{$^1$Centro Mixto CSIC-UPV/EHU,
             Aptdo. 1072, 20080 San Sebasti\'{a}n, Spain \\
         $^2$Donostia International Physics Center (DIPC),
             Aptdo. 1072, 20080 San Sebasti\'{a}n, Spain}
\date{\today}
\maketitle


\begin{abstract}
The interaction of a fast electron with a photonic crystal is
investigated by solving the Maxwell equations exactly for the
external field provided by the electron in the presence of the
crystal. The energy loss is obtained from the retarding force
exerted on the electron by the induced electric field. The
features of the energy loss spectra are shown to be related to the
photonic band structure of the crystal. Two different regimes are
discussed: for small lattice constants $a$ relative to the
wavelength of the associated electron excitations $\lambda$, an
effective medium theory can be used to describe the material;
however, for $a\sim\lambda$ the photonic band structure plays an
important role. Special attention is paid to the frequency gap
regions in the latter case.
\end{abstract}

]
\narrowtext


\section{Introduction}
\label{SecI}

Near-field spectroscopy can be performed using spatially-resolved
electron energy loss spectroscopy in scanning transmission
electron microscopes, where the electron beam provides an external
evanescent field with which to probe the sample within a spatial
range on the nanometer scale. In particular, the relatively
intense, low-energy part ($<50$ eV) of the loss spectrum can be
used with minimum sample damage to provide some insight on
plasmons and other collective excitations.
\cite{M82_1,C82_1,HM85_1,RH88_1} For the electron velocities
typically employed in microscopes (above half the speed of light)
and for samples that are homogeneous across distances of a few
nanometers, frequency-dependent dielectric functions are
sufficient to describe the materials that are involved and the
loss spectra reflect the geometry of the sample interfaces.
\cite{EP75_1,FE85_1,RZE92_1,DF94_1,GH98_1,G99_2}

The geometry becomes particularly important in photonic crystals,
where the periodic spatial modulation of the dielectric function
affects the propagation of radiation to the point of forbidding it
within band-gap energy regions. As a consequence, photonic
crystals are known to inhibit the spontaneous emission of light
within the band gap. \cite{Y87_1,JW90_1} They can also be used to
make omnidirectional dielectric mirrors that reflect light from
all directions without absorption \cite{HMT02_1} and wave guides
able to deflect light around sharp corners on the scale of the
wavelength. \cite{LCH98_1} These are applications of photonic
crystals that involve free external radiation, but equally
remarkable effects are expected to accompany evanescent fields
like those of external electrons.

In this work, we examine the energy loss spectra of electrons
moving near or inside photonic crystals over a wide range of
lattice parameters $a$. When $a$ is much smaller than the
wavelength of the radiation associated to a given energy loss,
$\lambda$, it is shown that the crystal can be described to some
extend by an effective dielectric function.
\cite{FMR93_1,FR93_1,BF95_1,MBF98_1} For larger lattice constants,
the photonic band structure becomes more complicated and this is
reflected in the loss spectra. The relation between the band
structure, the reflectance of photonic crystal slabs, and the
electron energy loss spectra is discussed in detail in what
follows.

In a previous development, Pendry and {Mart\'{\i}n-Moreno}
\cite{PM94_1} calculated the energy loss probability for electrons
moving near crystals made of either metallic spheres or metallic
cylinders in the $a<<\lambda$ limit using the transfer matrix
approach \cite{PM92_1} to solve Maxwell's equations. Their results
exhibit a complicated energy-loss structure even for relatively
dilute crystals , which has not been reproduced in the present
work. Therefore, in order to test the convergence of our method,
we have calculated the energy loss near the surface of a crystal
of dilute spheres and have found very good agreement between the
results derived from the theory described in this work and those
obtained from the analytical expression for isolated spheres.
\cite{GH98_1,G99_2}

In addition to producing energy loss, the interaction between the
electron and the crystal gives rise to the emission of the
so-called Smith-Purcell radiation. \cite{SP53_1} This effect has
already been discussed for one-dimensional \cite{G99_2} and
three-dimensional \cite{OY01_1,BGxx_1} crystals. Radiation
emission is one of the contributions to the total energy loss, and
in frequency regions where the crystal is transparent the
probability of these two must coincide. Examples of this are
offered below.

Here, the energy loss and photon emission probability are
calculated from the induced electric field, which is derived using
the reflection coefficients of the crystal, \cite{PM94_1} as
explained in Sec.\ \ref{SecII}. Results for the case of  small
lattice constants ($a<<\lambda$) are given in Sec.\ \ref{SecIII}
and for larger ones ($a\sim\lambda$) in Secs.\ \ref{SecIV} and
\ref{SecV}. The main conclusions are summarized in Sec.\
\ref{SecVI}. Gaussian atomic units (a.u., that is, $\hbar=m=e=1$)
will be used from now on, unless otherwise specified.


\section{Theory}
\label{SecII}

\subsection{The reflection coefficients of the crystal}
\label{SecIIa}

We shall consider crystals that are composed of a number of layers
perpendicular to the $z$ direction and extending from $z=0$
towards $z<0$. The host region outside the crystal will be assumed
to be described by a frequency-dependent dielectric function
$\epsilon_h(\omega)$ and a magnetic permeability $\mu_h(\omega)$.
Each crystal layer consists of the repetition of a given object
with certain two-dimensional translational symmetry that is shared
by all layers. The crystal will be characterized here by its
reflection coefficients, that is, the amplitudes of the reflected
plane wave components for a given incident wave.

In general, an external plane wave of frequency $\omega$ that
propagates near the crystal can be represented as
   \begin{eqnarray}
       \Eb^\pm_\Qb\,\exp(\ii \Kb^\pm_\Qb\cdot\rb),
   \nonumber
   \end{eqnarray}
where $\Qb=(Q_x, Q_y)$,
   \begin{eqnarray}
      \Kb^\pm_\Qb=(\Qb,\pm\ii\Gamma_Q),
   \nonumber
   \end{eqnarray}
   \begin{eqnarray}
      \Gamma_Q^2=Q^2-k_h^2,
   \nonumber
   \end{eqnarray}
and $k_h=(\omega/c) (\epsilon_h\mu_h)^{1/2}$ is the momentum of
the plane wave. The real part of $\Gamma_Q$ is chosen to be
positive, and the $+$ ($-$) sign in these expressions stands for a
wave moving towards $z>0$ ($z<0$). \cite{evanescentwaves}

The momentum $\Qb$ parallel to the surface has been singled out to
make explicit use of crystal symmetry: invoking momentum
conservation, a wave incident from the $z>0$ region with momentum
$\Kb^-_\Qb$ will only produce a discrete set of reflected waves of
momentum $\Kb^+_{\Qb+\Gb}$, where $\Gb$ runs over reciprocal
surface lattice vectors.

The transversal character of these waves (i.e., the fact that
$\Kb^\pm_\Qb\cdot\Eb^\pm_\Qb=0$) can be exploited to express the
electric field in terms of s and p components according to
   \begin{eqnarray}
      \Eb^\pm_\Qb=E^\pm_{\Qb,s}\,\eb_{\Qb,s}^\pm + E^\pm_{\Qb,p}\,\eb_{\Qb,p}^\pm,
   \nonumber
   \end{eqnarray}
where the vectors
   \begin{eqnarray}
      \eb_{\Qb,s}^\pm=\frac{1}{Q}(-Q_y,Q_x,0)
   \nonumber
   \end{eqnarray}
and
   \begin{eqnarray}
      \eb_{\Qb,p}^\pm=\frac{\ii}{k_h Q}(\pm\Gamma_Q Q_x,\pm\Gamma_Q Q_y,\ii Q^2)
   \nonumber
   \end{eqnarray}
satisfy the identities
$\eb_{\Qb,s}^\pm\cdot\eb_{\Qb,s}^\pm=\eb_{\Qb,p}^\pm\cdot\eb_{\Qb,p}^\pm=1$
and
$\eb_{\Qb,s}^\pm\cdot\eb_{\Qb,p}^\pm=\eb_{\Qb,s}^\pm\cdot\Kb^\pm_\Qb=\eb_{\Qb,p}^\pm\cdot\Kb^\pm_\Qb=0$.
Also, $\eb_{\Qb,s}^\pm$ is perpendicular to the plane defined by
$\Qb$ and the surface normal.

The amplitudes of the reflected waves depend linearly on the
amplitudes of s and p components of the incident wave, and the
coefficients of the linear relation between them are the
reflection coefficients $R_{\Qb\Gb}^{\sigma\sigma'}$, implicitly
defined by
   \begin{eqnarray}
      [E^+_{\Qb+\Gb,\sigma}]^r = \sum_{\sigma'}
                             R_{\Qb\Gb}^{\sigma\sigma'}
                             [E^-_{\Qb,\sigma'}]^i,
   \label{e1}
   \end{eqnarray}
where $\sigma$ and $\sigma'$ run over polarization directions s
and p, and the super-indices $r$ and $i$ stand for reflected and
incident components, respectively.

We have used the layer KKR method to calculate the reflection
coefficients $R_{\Qb\Gb}^{\sigma\sigma'}$ both for $a<<\lambda$
and for $a\sim\lambda$. In the layer KKR method, the transmission
and reflection coefficients are calculated exactly for each single
layer using multiple scattering in a basis set of multipoles
centered around each object of the layer. Scattering among layers
is then expressed in terms of those coefficients. The maximum
multipole order $\lmax$ and the number of reflected and
transmitted beams (i.e., the number of $\Gb$ vectors) are the only
convergence parameters, which have been tested in all calculated
results shown below. Stefanou {\it et al.} \cite{SYM98_1}
developed this method for spheres and we have extended it to be
able to deal with non-spherical objects and arbitrary values of
$\lmax$. Further details of the method will be given elsewhere.
\cite{Gxx_1}

\subsection{The field of the electron}
\label{SecIIb}

We shall consider an electron moving with constant velocity $v$
along a trajectory parallel to the crystal surface and described
by $\rb_t=(v t, y_0, z_0)$, with $z_0>0$, so that the electron
moves in the host medium described by $\epsilon_h$ and $\mu_h$.
Neglecting the crystal for the moment, the electron electric field
can be expressed in frequency space $\omega$ in terms of vector
and scalar potentials as
   \begin{eqnarray}
      \Eb_0=\frac{\ii\omega}{c} \Ab_0-\nabla\phi_0.
   \nonumber
   \end{eqnarray}
In the Lorentz gauge, Maxwell's equations can be recast as
   \begin{eqnarray}
      (\nabla^2+k_h^2)\phi_0=-\frac{4\pi}{\epsilon_h}\rho
   \nonumber
   \end{eqnarray}
and
   \begin{eqnarray}
      (\nabla^2+k_h^2)\Ab_0=-\frac{4\pi\mu_h}{c}\jb,
   \nonumber
   \end{eqnarray}
where $\rho(\rb,\omega)$ is the electron charge density, and
$\jb(\rb,\omega)=(v \rho, 0, 0)$ is its charge current.

Using the relation
   \begin{eqnarray}
      (\nabla^2+k_h^2)\int \frac{d\pb}{2\pi^2}
           \frac{\ee^{\ii \pb\cdot\rb}}
                {p^2-k_h^2-\ii 0^+}=-4\pi\delta(\rb)
   \nonumber
   \end{eqnarray}
and noticing that $\rho(\rb,t)=-\delta[\rb-(v t, y_0, z_0)]$, the
electric field is found to be
   \begin{eqnarray}
      \Eb_0(\rb,\omega) &=& [\frac{\nabla}{\epsilon_h}
                         -\frac{\ii\omega v \mu_h}{c^2} \xh]
      \nonumber \\ && \times \,
                       \int \frac{d\pb}{2\pi^2}
                       \int dt \, \ee^{\ii\omega t}
                       \frac{\ee^{\ii \pb\cdot[\rb-(v t, y_0, z_0)]}}
                {p^2-k_h^2-\ii 0^+},
   \nonumber
   \end{eqnarray}
where the time integral represents the inverse Fourier transform
that permits obtaining $\rho(\rb,\omega)$ in terms of
$\rho(\rb,t)$. The above integral can be reduced to
   \begin{eqnarray}
      \Eb_0(\rb,\omega)=\int dQ_y \,
                        \ee^{\ii \Kb^\pm_\Qb\cdot[\rb-(0,y_0,z_0)])}
                        \Eb^\pm_\Qb,
   \label{e2}
   \end{eqnarray}
where $\Qb=(\omega/v,Q_y)$ and
   \begin{eqnarray}
      \Eb^\pm_\Qb=\frac{\ii}{\Gamma_Q}
                (\frac{\Kb_\Qb^\pm}{v \epsilon_h}-\frac{\omega\mu_h}{c^2} \xh).
   \nonumber
   \end{eqnarray}
The $+$ ($-$) sign must be used in these expressions when $z>z_0$
($z<z_0$), so that the integrand of Eq.\ (\ref{e2}) represents a
plane wave that propagates towards positive (negative) $z$'s. When
the electron is moving in vacuum, $\Gamma_Q$ is real and the waves
in the integrand of Eq.\ (\ref{e2}) are evanescent.
\cite{evanescentwaves} However, when $k_h$ is real and larger than
$\omega/v$, some of those waves describe Cherenkov radiation that
propagates without attenuation; this is the case of electrons that
travel faster than light in the medium.

\subsection{The field induced by interaction of the electron and the crystal}
\label{SecIIc}

The decomposition of these plane waves into s and p components is
readily found to be
   \begin{eqnarray}
      [E^\pm_{\Qb,s}]^i=\frac{\ii Q_y\omega\mu_h}{Q\Gamma_Q c^2}\ee^{-\ii \Kb^\pm_\Qb\cdot(0,y_0,z_0)}
   \nonumber
   \end{eqnarray}
and
   \begin{eqnarray}
      [E^\pm_{\Qb,p}]^i=\pm \frac{k_h}{v Q\epsilon_h}\ee^{-\ii \Kb^\pm_\Qb\cdot(0,y_0,z_0)}.
   \nonumber
   \end{eqnarray}
Each of the incident plane waves [i.e., each value of $Q_y$ in
Eq.\ (\ref{e2})] gives rise to a set of reflected waves whose
amplitudes are obtained from Eq.\ (\ref{e1}). Therefore, the
electric field in the region near the ion can be constructed as
the sum of $\Eb_0$ and the reflected field,
   \begin{eqnarray}
      \Eb(\rb,\omega)=\Eb_0(\rb,\omega)+\Eb_r(\rb,\omega),
   \label{e3}
   \end{eqnarray}
where
   \begin{eqnarray}
      &&\Eb_r(\rb,\omega)=
   \nonumber \\
      &&\sum_{\Gb,\sigma\sigma'} \int dQ_y \,\,
          \ee^{\ii \Kb^+_{\Qb+\Gb}\cdot\rb} \,
          R_{\Qb\Gb}^{\sigma\sigma'} \, [E^-_{\Qb,\sigma'}]^i \,\,
          \eb_{\Qb+\Gb,\sigma}^+,
   \label{e4}
   \end{eqnarray}
and the integral over $Q_y$ has been copied directly from Eq.\
(\ref{e2}) in virtue of the linearity of Maxwell's equations.

\subsection{Electron energy loss in front of the crystal}
\label{SecIId}

The electron energy loss can be calculated from the retarded force
exerted by the induced part of the electric field $\Eb^{\rm ind}$
acting back on the electron. Integrating this force along the
trajectory and dividing by the total path length $L$, one finds
   \begin{eqnarray}
      \frac {\Delta E}{\Delta x}
               = \frac{v}{L}\int dt \; E_x^{\rm ind}(\rb_t,t)
               = \int_0^\infty \omega d\omega \; P(\omega),
   \nonumber
   \end{eqnarray}
where
   \begin{eqnarray}
      P(\omega) = \frac{v}{\pi\omega L} \int dt \,
                  {\rm Re} \{ \ee^{-\ii\omega t}
              E_x^{\rm ind}(\rb_t,\omega) \}
   \label{e5}
   \end{eqnarray}
is the loss probability per unit of path length.

For an electron moving parallel to a crystal surface, Eq.\
(\ref{e3}) permits separating the loss probability as
   \begin{eqnarray}
      P=P_0+P_r,
   \nonumber
   \end{eqnarray}
where \cite{GH98_1}
   \begin{eqnarray}
      P_0(\omega) &=& \frac{1}{\pi v^2}
         {\rm Im} \{(\frac{v^2}{c^2}\mu_h-\frac{1}{\epsilon_h})
                            \log[\frac{q_c^2-k_h^2}{(\omega/v)^2-k_h^2}]\}
   \nonumber
   \end{eqnarray}
corresponds to the contribution of $\Eb_0$ in the absence of the
crystal. Here, $q_c$ is a momentum cut-off related to energy
conservation. This contribution vanishes in vacuum. However, it
gives rise to Cherenkov losses when $\epsilon_h$ and $\mu_h$ are
real, in which case \cite{J75_1,GH98_1}
$P_0=|\mu_h|[1/c^2-1/(v^2\epsilon_h\mu_h)]$, subject to the
Cherenkov condition $v^2\epsilon_h\mu_h>c^2$.

The remaining contribution $P_r$ is due to the retarding force
exerted by the reflected field $\Eb_r$. Inserting Eq.\ (\ref{e4})
into Eq.\ (\ref{e5}), the time $t$ and the lateral impact
parameter $y_0$ appear only through the factor $\exp\{\ii(G_x v t
+ G_y y_0)\}$, so that time integration eliminates all vectors
$\Gb$ with $G_x\neq 0$ from the sum in Eq.\ (\ref{e4}).
Furthermore, averaging over $y_0$ leaves only the $G_y=0$ term and
one obtains
   \begin{eqnarray}
      P_r(\omega) &=& \frac{1}{\pi v^2}
            \int \frac{\, dQ_y}{Q^2} \,\,
          \,\,
   \label{e6} \\ &\times&
          {\rm Im} \{[R_{\Qb0}^{pp}
                      + \epsilon_h\mu_h (\frac{Q_y v}{\Gamma_Q c})^2 R_{\Qb0}^{ss}
   \nonumber \\
           &&  \;\;\;\;  + \frac{\ii Q_y k_h v}{\omega\Gamma_Q}
                             (R_{\Qb 0}^{sp} - R_{\Qb0}^{ps})
          ]  \frac{\Gamma_Q\ee^{-2\Gamma_Q z_0}}{\epsilon_h} \}.
   \nonumber
   \end{eqnarray}
The last term inside the square bracket of this expression gives
no contribution when the electron trajectory is contained in a
plane of specular symmetry of the crystal surface.

As a test one can apply Eq.\ (\ref{e6}) to an electron moving in
vacuum ($\epsilon_h=\mu_h=1$) at a distance $z_0$ from the surface
of a non-magnetic medium described by $\epsilon$, in which case
$R_{\Qb0}^{sp}=R_{\Qb0}^{ps}=0$. Then, Eq.\ (\ref{e6}) reduces to
   \begin{eqnarray}
      P(\omega) &=& \frac{2}{\pi v^2}
            \int_0^\infty \frac{\Gamma_Q\, dQ_y}{Q^2} \,\,
          \ee^{-2\Gamma_Q z_0} \,\,
   \nonumber \\ &\times& \,
          {\rm Im} \{
          \frac{\epsilon\Gamma_Q-\Gamma'_Q}{\epsilon\Gamma_Q+\Gamma'_Q}
          +(\frac{Q_y v}{\Gamma_Q c})^2
          \frac{\Gamma_Q-\Gamma'_Q}{\Gamma_Q+\Gamma'_Q}
          \},
   \nonumber
   \end{eqnarray}
where $\Gamma'_Q=\sqrt{Q^2-\epsilon\omega^2/c^2}$,
$\Gamma_Q=\sqrt{Q^2-\omega^2/c^2}$, and the reflection
coefficients of Eq.\ (\ref{e6}) have been taken from Fresnel's
equations. \cite{J75_1} In the non-relativistic limit,
$\Gamma_Q=\Gamma'_Q=Q$, so that the contribution of the reflection
of s waves vanishes and the loss probability becomes proportional
to ${\rm Im}\{-1/(\epsilon+1)\}$.


\section{Energy loss for small lattice constants}
\label{SecIII}

When the crystal lattice constant is much smaller than the
wavelength corresponding to a given frequency component $\omega$,
the details of the crystal lattice cannot be resolved by the
electron, so that the medium behaves like a uniform material,
characterized by an effective dielectric constant and magnetic
permeability. For instance, a dilute system of spheres can be
regarded as a set of interacting dipoles, which leads to the
well-known Maxwell-Garnett formula. \cite{M04_1} For more compact
systems the details of the microscopic structure becomes relevant
via important multipolar interactions.

These are the cases considered in Fig.\ \ref{Figd} for a crystal
composed of small aluminum spheres that are disposed in a simple
cubic lattice of constant $a=5$ nm. The solid curves of Fig.\
\ref{Figd}(a) show the loss probability for an electron moving
with velocity $v=0.4 c$ parallel to the surface of a slab made up
of six $(100)$ layers of such crystal. The lost probability has
been calculated by means of Eq.\ (\ref{e6}) for two different
filling fractions $f$ of the aluminum ($6.5 \%$ and $30 \%$,
respectively). The dielectric function of aluminum has been
approximated by a Drude expression
$\epsilon(\omega)=1-\omega_p^2/\omega(\omega+\ii\eta)$ with
$\omega_p=15$ eV and $\eta=1$ eV. The wavelengths considered in
the figure lie in the range $\lambda=2\pi c/\omega=88.6-310$ nm,
so that we are in the $\lambda>>a$ limit. The maximum orbital
quantum number used to achieve convergence on the figure is
$\lmax=6$.

Crystalline effects come from the interaction among spheres, and a
method to determine their relative role consists in comparing
these results with a calculation in which that interaction is
suppressed. This is what the broken curves stand for. It is very
clear that the interaction among spheres is almost negligible for
$f=6.5\%$, whereas it becomes very important at $f=30\%$. In the
latter case, the sphere surfaces are closer to each other and
their mutual electromagnetic coupling becomes relevant. The
non-interacting case shows a peak at around $\omega_1=8.7$ eV that
corresponds to the dipole Mie resonance in the small-sphere limit,
given by the expression $\epsilon(\omega_1)=-2$ (i.e.,
$\omega_1=\omega_p/\sqrt{3}$). The main effect of the
sphere-sphere interaction (solid curve for $f=30\%$) consists in
splitting this peak, in a similar way as splitting of degenerate
levels (the Mie resonances here) occurs in atomic bonding.

This is actually observed in the projected photonic band structure
of this crystal, represented in Fig.\ \ref{Figc} for $f=30 \%$.
The structure is dominated by nearly flat bands corresponding to
localized excitations near the Mie resonances of the small
isolated spheres, $\omega_l=\omega_p\sqrt{l/(l+1)}$, which lie in
the 8.7-10.6 eV range (see the labels on the right hand side of
the figure for the dipole and quadrupole Mie modes). The
interaction between spheres gives rise to a complex structure that
encompasses regions of relative transparency outside that energy
range. The evanescent plane wave components of the external
electron field are subject to the condition $\omega=\Qb\cdot\vb$,
which defines the straight lines shown in the figure. The loss
spectrum can then be understood as originating from both
absorption and direct coupling to propagating modes of the
crystal. This last effect is clearly seen as a bump in the loss
probability near 6.7 eV, connected to the crossing of the noted
straight lines with a low-energy region of allowed propagating
modes.

The dashed curves of Fig.\ \ref{Figd}(a) (uncoupled spheres) have
been obtained in two different ways. The first one consists in
calculating the reflection coefficients that appear in Eq.\
(\ref{e6}) using the layer KKR method but neglecting the
interaction among spheres within each layer and also the
interaction among different layers. A second procedure consists in
making use of the analytical expression for the energy loss
probability of an electron near an isolated sphere (see Refs.\
[\onlinecite{FE85_1}] and [\onlinecite{GH98_1,G99_2}] for
non-relativistic and fully-relativistic formulas, respectively)
and summing over all impact parameters of the electron trajectory
with respect to the spheres of the crystal. The results coming out
of these two very different procedures cannot be distinguished on
the scale of the figure, and this is a strong indication of the
degree of convergence of our numerical calculations with respect
to the number of plane waves used in the layer KKR method.

These results do not support, however, previous calculations by
Pendry and {Mart\'{\i}n-Moreno} \cite{PM94_1} for this exact
system, where the number of features and their energy position for
the filling fractions under consideration differ from ours. A
possible lack of convergence of the transfer matrix method used by
the authors for this three-dimensional system of metallic
scatterers might be the reason of this discrepancy. In particular,
some of their low-energy features around 6.5 eV could originate in
the modes of the wedge associated to their space discretization
procedure (resonances near that energy have also been found for
the $90^\circ$ aluminum wedge \cite{GH98_1}).

The example offered in Fig.\ \ref{Figd} illustrates what happens
in the $a<<\lambda$ limit, where it is reasonable to define an
effective dielectric function $\epsilon_{\rm eff}$ for the
infinite crystal and to compare the results of the detailed, exact
calculation with those obtained for an electron moving parallel to
a homogeneous surface of a material described by such a dielectric
function. Several recipes for defining $\epsilon_{\rm eff}$ for
granular materials have been given in the literature, ranging from
simple effective medium theories like Maxwell-Garnett's
\cite{M04_1} to more elaborate ones that take into account the
actual shape of the constituents, both for disordered composites
\cite{LWA80_1,CKK90_1,BF95_1} and for crystals.
\cite{LWA80_1,BD92_1,MP95_1,MZT00_1,SSM02_1} Among the latter, one
finds extensions of the Maxwell-Garnett theory that go beyond
dipolar interactions, \cite{LWA80_1,BF95_1} spectral
representations, \cite {BD92_1,MZT00_1} or direct derivation of
the light dispersion relation. \cite{MP95_1} Here, we have used a
different method that consists in finding the dielectric function
$\epsilon_{\rm eff}$ of the equivalent homogeneous medium that has
the same reflectance as the crystal surface; \cite{Gxx_1} the
reflectance calculated for the crystal is well reproduced by
Fresnel equations with a single value of $\epsilon_{\rm
eff}(\omega)$ for each frequency $\omega$ within a $2\%$ under the
present conditions. The results are represented by the solid
curves of Fig.\ \ref{Figd}(b), where they are compared with
Maxwell-Garnett's theory (broken curves). For $f=6.5\%$ these two
models are nearly identical, as Maxwell-Garnett formula gives the
dipole of the small isolated sphere correctly. However, for $f=30
\%$ the deviation between both models is significant, since higher
momenta are involved in the interaction between neighboring
spheres . The magnitude actually represented in the figure is the
surface loss function, ${\rm Im}\{-1/(1+\epsilon_{\rm eff})\}$,
which is directly comparable to the loss probability of Fig.\
\ref{Figd}(a). The agreement in the position of the peaks between
the detailed energy loss calculation and our effective medium
theory is reasonably good, indicating that an effective
homogeneous medium describes the solid appropriately within this
energy range at $v=0.4 c$; however, the relative weight of the
features for $f=30 \%$ changes completely at lower velocities
(dotted curve, $v=0.06 c=8.2$ a.u.).

Another example of a crystal that can be represented by an
effective dielectric constant is given in Fig.\ \ref{Fige},
consisting of a simple cubic lattice with the same parameters as
in Fig.\ \ref{Figd}, except that finite cylinders have been used
instead of spheres. As a result, the medium is strongly
anisotropic and characterized by different bulk plasmon modes when
the electric field is directed parallel or perpendicular to the
cylinders [see Fig.\ \ref{Fige}(c)-(d)]. The loss probability
[Fig.\ \ref{Fige}(a)] is shown to share most of the features of
the surface loss function for anisotropic media [Fig.\
\ref{Fige}(b)], except for the 7.8 eV peak in the loss spectrum
with $f=30 \%$, which might be connected to the proximity effect
when the electron starts sensing the non-uniform character of the
surface via evanescent waves.


\section{Energy loss for lattice constants comparable to the wavelength}
\label{SecIV}

For lattice constants comparable to the wavelength associated to
the energy losses under consideration, one can no longer define
the effective dielectric constant of an equivalent homogeneous
medium. Then, it is useful to relate the loss spectra directly to
the photonic band structure. This has been done in Fig.\
\ref{Figb} for a crystal of aluminum spheres immersed in a
dielectric with $\epsilon=3$. The features of the loss spectra
[Fig.\ \ref{Figb}(c)] are strongly correlated with the band
structure [Fig.\ \ref{Figb}(a)-(b)]. In particular, the
pseudo-gaps near 3.2 eV are translated into a dip in the loss
probability.

Using electrons to analyze the crystal permits exploring regions
of the band structure that are not accessible to external light,
but that can be reached via the evanescent waves contained in the
perturbing field of the electrons. This is the case of Fig.\
\ref{Figb} for $v/c=0.5$ [solid curve in Fig.\ \ref{Figb}(c)],
well below the Cherenkov threshold $v/c=1/\sqrt{3}$, that permits
to observe how the peak in the loss spectrum near 2.7 eV results
from the coupling with propagating modes of the crystal as a
result of the intersection of the external field components
[straight solid lines in Fig.\ \ref{Figb}(a)] and the allowed
regions of propagation (shaded areas).

At larger velocities [$v/c=0.75$, dashed curve in Fig.\
\ref{Figb}(c)], Cherenkov radiation would be produced in the
absence of the crystal, that would result in a loss probability
independent of $\omega$ (see little arrow near the horizontal axis
of the figure). This is clearly seen in the big overlap of the
external field with the regions of propagation within the crystal
inside the low-energy region results in an enhanced loss
probability, as compared to $v/c=0.5$. However, the presence of
the crystal modulates the loss spectrum.

Part of the energy lost by the electron must be converted into the
so-called Smith-Purcell radiation. \cite{SP53_1,G99_2} In
particular, when transparent materials are used to build the
crystals, the light emission probability must coincide with the
energy-loss probability. For a crystal of finite thickness, like
the one considered in Fig.\ \ref{Figf}, consisting of 8
(111)-layers of an inverted Si opal ($\epsilon=11.9$) with a
filling fraction of $67 \%$, part of this emission occurs towards
the side of the crystal opposite to the electron trajectory (here,
the electron is taken to be moving parallel to the crystal
surface). The intensity of the emitted light [Fig.\ \ref{Figf}(c)]
has been calculated from the integral of the Poynting vector far
from the crystal (see Ref.\ [\onlinecite{BGxx_1}] for more
details), and it presents a strong dip near $a/\lambda\approx
0.75$, where $a=1220$ nm is the lattice constant. The region of
emission depletion is actually contained within a full band-gap of
the infinite crystal [see Fig.\ \ref{Figf}(a)]; this is also seen
in the transmission of light incident on the crystal both normal
to the surface [dashed curve in Fig.\ \ref{Figf}(b)] or with an
angle corresponding to the Cherenkov radiation produced by the
electron in the plane defined by the trajectory and the surface
normal (solid curve).


\section{Energy loss in the bulk of a crystal}
\label{SecV}

So far we have considered an electron moving in front of the
surface of a photonic crystal. When the electron is moving in the
bulk of an infinite crystal, the electric field can also be
written in terms of reflectance coefficients if the trajectory is
contained in a low-index plane that does not intersect any of the
crystal objects. The reflectance in question is that of the lower
and upper semi-infinite crystals into which the entire crystal is
divided by the noted plane. The corresponding reflectance matrices
will be denoted $R_1$ and $R_2$, respectively. They contain the
amplitudes of reflected beams, labeled by vectors $\Gb$ of the 2D
reciprocal lattice of the plane under consideration. In
particular, for a fixed value of $\Qb$,
$R_{1,\Gb\Gb'}^{\sigma\sigma'}$ is given by
$R_{\Qb+\Gb',\Gb-\Gb'}^{\sigma\sigma'}$, as defined by Eq.\
(\ref{e1}). Similarly, $R_{2,\Gb\Gb'}^{\sigma\sigma'}$ connects
$\Kb_{\Qb+\Gb}^+$ wave components with waves reflected from the
upper surface with momentum $\Kb_{\Qb+\Gb'}^-$.

This is represented schematically in Fig.\ \ref{Figa}, where the
reflectance matrices of the lower and upper semi-infinite crystals
are $R_1$ and $R_2$, respectively, and upwards (downwards) arrows
represent waves of momentum $\Kb_{\Qb+\Gb}^+$ ($\Kb_{\Qb+\Gb}^-$).
The upper and lower surfaces have been separated in the figure for
the sake of clarity, but they will be considered to lie on the
same plane (a plane that contains the electron trajectory) in what
follows.

Using Eq.\ (\ref{e5}) for the loss probability and a
straightforward extension of Eq.\ (\ref{e4}) for the reflected
electric field including all terms of Fig.\ \ref{Figa}, the loss
probability per unit of path length averaged over all impact
parameters parallel to the crystal surfaces is found to be
   \begin{eqnarray}
      P_r(\omega) &=& \frac{1}{\pi v^2}
            \int \frac{\, dQ_y}{Q^2} \,\,
          \,\,
   \label{e7} \\ &\times&
          {\rm Im} \{[A_{00}^{pp}+B_{00}^{pp}-C_{00}^{pp}-D_{00}^{pp}
   \nonumber \\ &&
                         + \epsilon_h\mu_h (\frac{Q_y v}{\Gamma_Q c})^2
                         (A_{00}^{ss}+B_{00}^{ss}+C_{00}^{ss}+D_{00}^{ss})
   \nonumber \\
           &&  \;\;\;\;  + \frac{\ii Q_y k_h v}{\omega\Gamma_Q}
                             (A_{00}^{sp}-B_{00}^{sp}+C_{00}^{sp}-D_{00}^{sp}
   \nonumber \\ && \;\;\;\;\;\;\;\;\;\;\;
                              - A_{00}^{ps}+B_{00}^{ps}+C_{00}^{ps}-D_{00}^{ps})
          ]  \frac{\Gamma_Q}{\epsilon_h} \},
   \nonumber
   \end{eqnarray}
where the matrices $A_{\Gb\Gb'}^{\sigma\sigma'}$,
$B_{\Gb\Gb'}^{\sigma\sigma'}$, $C_{\Gb\Gb'}^{\sigma\sigma'}$, and
$D_{\Gb\Gb'}^{\sigma\sigma'}$ are defined in Fig.\ \ref{Figa}, and
only the elements $\Gb=\Gb'=0$ enters this expression.

The above formula has been applied to calculate the energy loss
probability for an electron moving in between two parallel crystal
slabs of 32 layers each surrounded by Si, as shown in Fig.\
\ref{Fig7}. The crystal parameters are the same as in Fig.\
\ref{Figf}(c), so that an absolute band gap shows up near
$a/\lambda=0.75$. The electron energy (100 keV) is large enough to
produce Cherenkov light, which must be confined in between the two
crystals for wavelengths lying in the gap region. This is actually
observed as a pronounced dip, connected to the fact that no
electromagnetic modes can be created that escape through the
crystals, except for a small transmission due to their finite
thickness. However, some radiation can escape though modes that
are trapped in the slab formed by the two crystals, so that the
loss probability does not actually reach a zero value in the dip.


\section{Concluding remarks}
\label{SecVI}

The energy loss of fast electrons interacting with photonic
crystals has been calculated for dielectric, metallic, and
metalo-dielectric systems.

\begin{figure}
\centerline{\scalebox{1.00}{\includegraphics{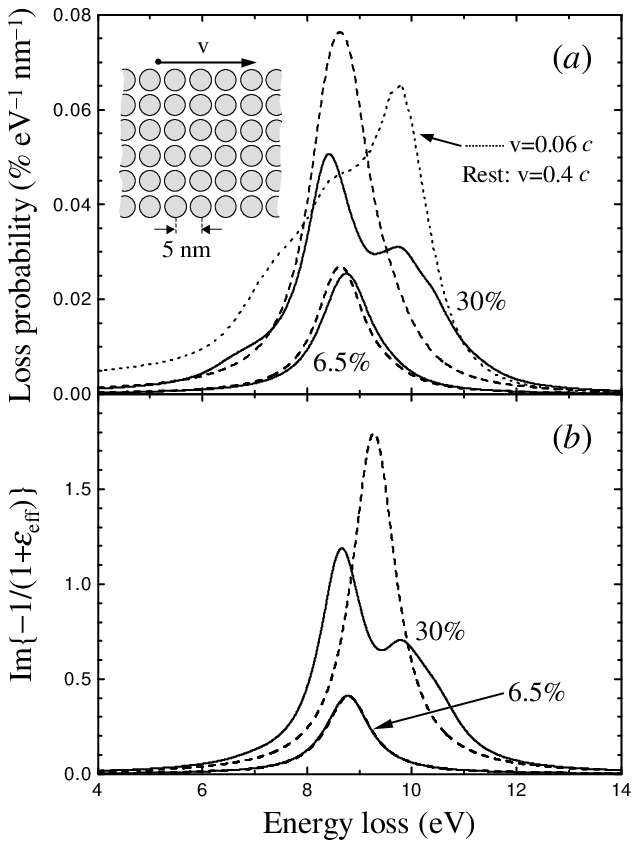}}}
\caption{{\bf (a)} Energy loss spectra for an electron moving
parallel to the $[100]$ direction of the $(100)$ surface of a
simple cubic crystal made up of six layers of aluminum spheres in
vacuum with lattice constant 5 nm and two different filling
fractions (see labels). The electron is moving at a distance of 1
nm from the sphere surfaces with a velocity $v=0.4 c$ (solid and
dashed curves) and $v=0.06 c$ (dotted curve). The dashed curves
are obtained by neglecting the interaction among spheres, whereas
the solid curves and the dotted curve correspond to the full
solution of Maxwell's equations. {\bf (b)} Surface loss function
for the same crystal as in (a) using the effective dielectric
function obtained from the Maxwell-Garnett expression (broken
curves) and from the multiple-scattering method explained in the
text (solid curves).} \label{Figd}
\end{figure}

\begin{figure}
\centerline{\scalebox{1.30}{\includegraphics{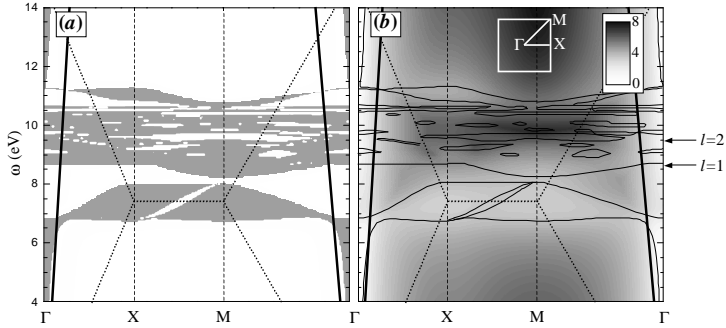}}}
\caption{Photonic band structure of one of the crystals considered
in Fig.\ \ref{Figd}, consisting of a simple cubic lattice of
aluminum spheres in vacuum with a lattice constant of 5 nm and a
filling fraction of $30 \%$. The figure shows the band structure
projected on the $(100)$ surface. The horizontal axis represents
the excursion along the points indicated in the inset within the
parallel momentum plane. The expression
$\epsilon(\omega)=1-\omega_p^2/\omega^2$ has been used for the
aluminum dielectric constant with $\omega_p=15$ eV. When the
damping is taken as $\eta\rightarrow 0^+$, one obtains regions of
allowed electromagnetic propagation, that is, combinations of the
energy and the parallel momentum components for which some
eigenstates have a real perpendicular momentum. They define the
shaded region in {\bf (a)}. However, for a realistic value of the
damping parameter ($\eta=1$ eV), all perpendicular momentum
components are complex and they represent evanescent waves within
the crystal. The contour plot in figure {\bf (b)} represents the
lowest value of the imaginary part of the perpendicular momentum
multiplied by the lattice constant (i.e., the minimum of ${\rm
Im}\{k_z a\}$) for all eigenstates with a given parallel momentum.
The boundaries of the shaded areas of (a) are shown in (b) as
continuous curves.} \label{Figc}
\end{figure}

\begin{figure}
\centerline{\scalebox{1.00}{\includegraphics{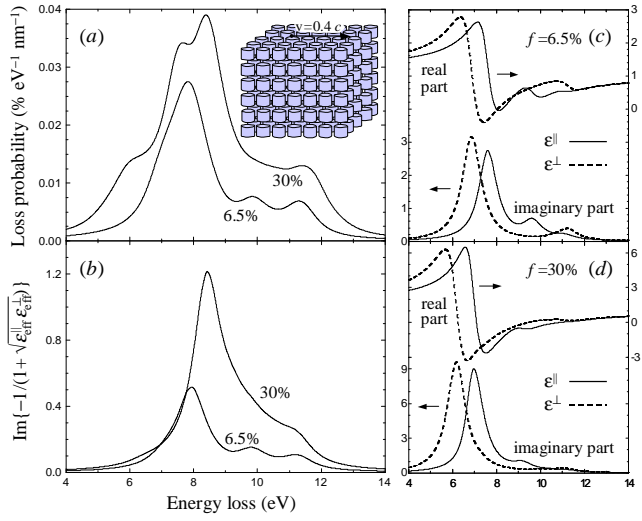}}}
\caption{{\bf (a)} Energy loss spectra for an electron moving
parallel to the $[100]$ direction of the $(100)$ surface of a
simple cubic crystal made up of six layers of aluminum cylinders
in vacuum with lattice constant 5 nm and two different filling
fractions (see labels). The height of the cylinders is equal to
the diameter in all cases. The electron is moving at a distance of
1 nm from the cylinder surfaces with a velocity $v=0.4 c$. {\bf
(b)} Surface loss function for the same crystal as in (a) using
the effective dielectric function obtained from the
multiple-scattering method explained in the text. {\bf (c)}
Anisotropic dielectric function for the crystal considered in (a)
with a filling fraction of the aluminum of $6.5 \%$. {\bf (d)}
Same as (c), for a filling fraction of $30 \%$.} \label{Fige}
\end{figure}

When the wavelength of the radiation associated to the energy loss
is much larger than the lattice constant, the crystal can be
regarded as a continuous medium, characterized by an effective
dielectric function, as it has been shown for crystals of small
aluminum spheres in Fig.\ \ref{Figd} and small finite cylinders in
Fig.\ \ref{Fige}. The effective medium is highly anisotropic in
the latter case. The effective dielectric function has been
calculated in both cases from the reflectance coefficient of the
crystal, and this has been shown to contain most of the
information needed to understand the calculated energy-loss
spectra.

\begin{figure}
\centerline{\scalebox{1.00}{\includegraphics{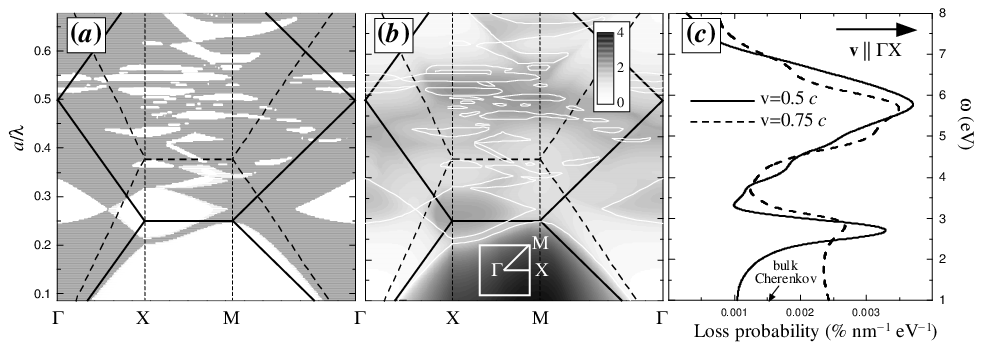}}}
\caption{{\bf (a)-(b)} Photonic band structure projected on the
(100) surface and represented as in Fig.\ \ref{Figc}(a)-(b) for an
fcc lattice of aluminum spheres surrounded by a medium of
dielectric function equal to 3. The filling fraction of the
spheres is $20 \%$ and the lattice constant is $a=105.2$ nm,
comparable to the wavelength $\lambda$ (see scale on the left hand
side of the figure). {\bf (c)} Electron energy loss spectra for an
electron moving with velocity $v=0.5 c$ (solid curve) and $v=0.75
c$ (broken curve) at a distance of 3.2 nm from the surface of the
outer-most spheres of a semi-infinite crystal like that considered
in (b). The trajectory is directed along the [100] direction
($\Gamma X$). The frequency and parallel momentum of the plane
wave components of the external electron field are related by the
condition $\omega=\Qb\cdot\vb$, which defines the straight lines
shown in (a) and (b).} \label{Figb}
\end{figure}

\begin{figure}
\centerline{\scalebox{0.36}{\includegraphics{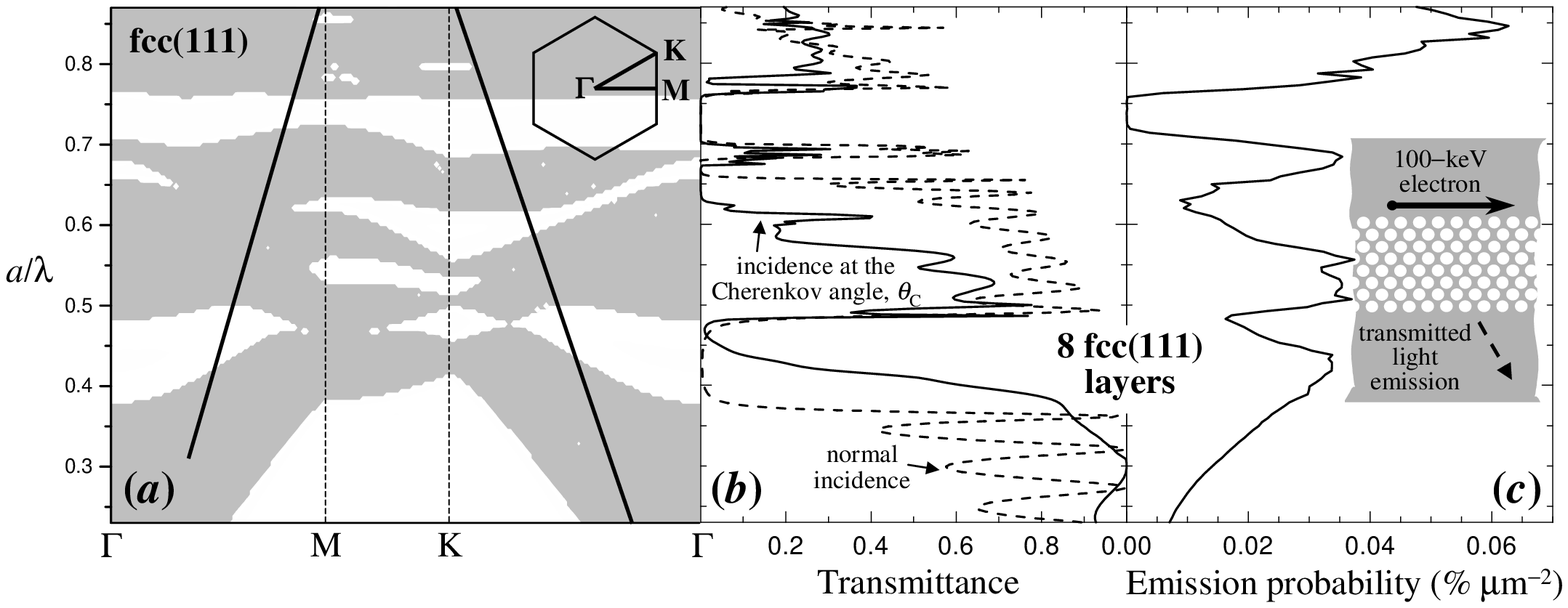}}}
\caption{{\bf (a)} Photonic band structure projected on the (111)
surface for an fcc lattice of air spherical voids in Si
($\epsilon=11.9$). The radius of the spheres is 0.342 times the
lattice constant, $a$. {\bf (b)} Dashed curve: transmittance of
light incident normal to 8 layers of the crystal considered in
(a). Solid curve: transmittance for an angle of incidence
corresponding to the Cherenkov angle $\theta_c=68.1^\circ$ for a
100-keV electron moving parallel to the same crystal. The crystal
is surrounded by Si on both surfaces . {\bf (c)} Probability of
emitting light on the opposite side of the crystal with respect to
the electron trajectory. The distance from the latter to the
sphere surfaces is 876 nm and the lattice constant is $a=1220$ nm.
The probability is given per unit of path length in microns and
per unit of emitted-photon wavelength also in microns. The region
of frequency and momentum where the field of the electron inside a
homogeneous infinite Si medium takes non-zero values is
represented by thick solid lines in (a).} \label{Figf}
\end{figure}

\begin{figure}
\centerline{\scalebox{0.36}{\includegraphics{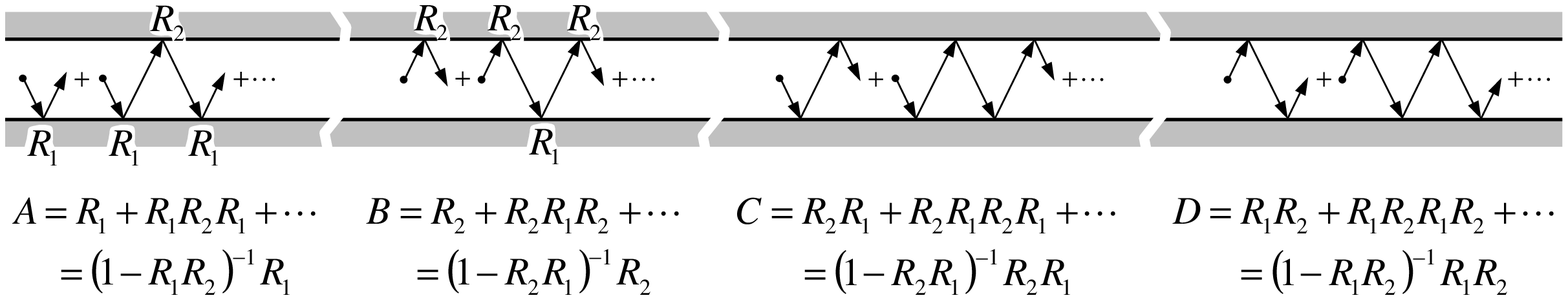}}}
\caption{Schematic representation of the construction of the
electric field in the bulk of a photonic crystal in terms of the
reflectance of two semi-infinite crystals and definition of the
matrices $A$, $B$, $C$, and $D$ used in Eq.\ (\ref{e7}). See Sec.\
\ref{SecV} for more details.} \label{Figa}
\end{figure}

\begin{figure}
\centerline{\scalebox{0.36}{\includegraphics{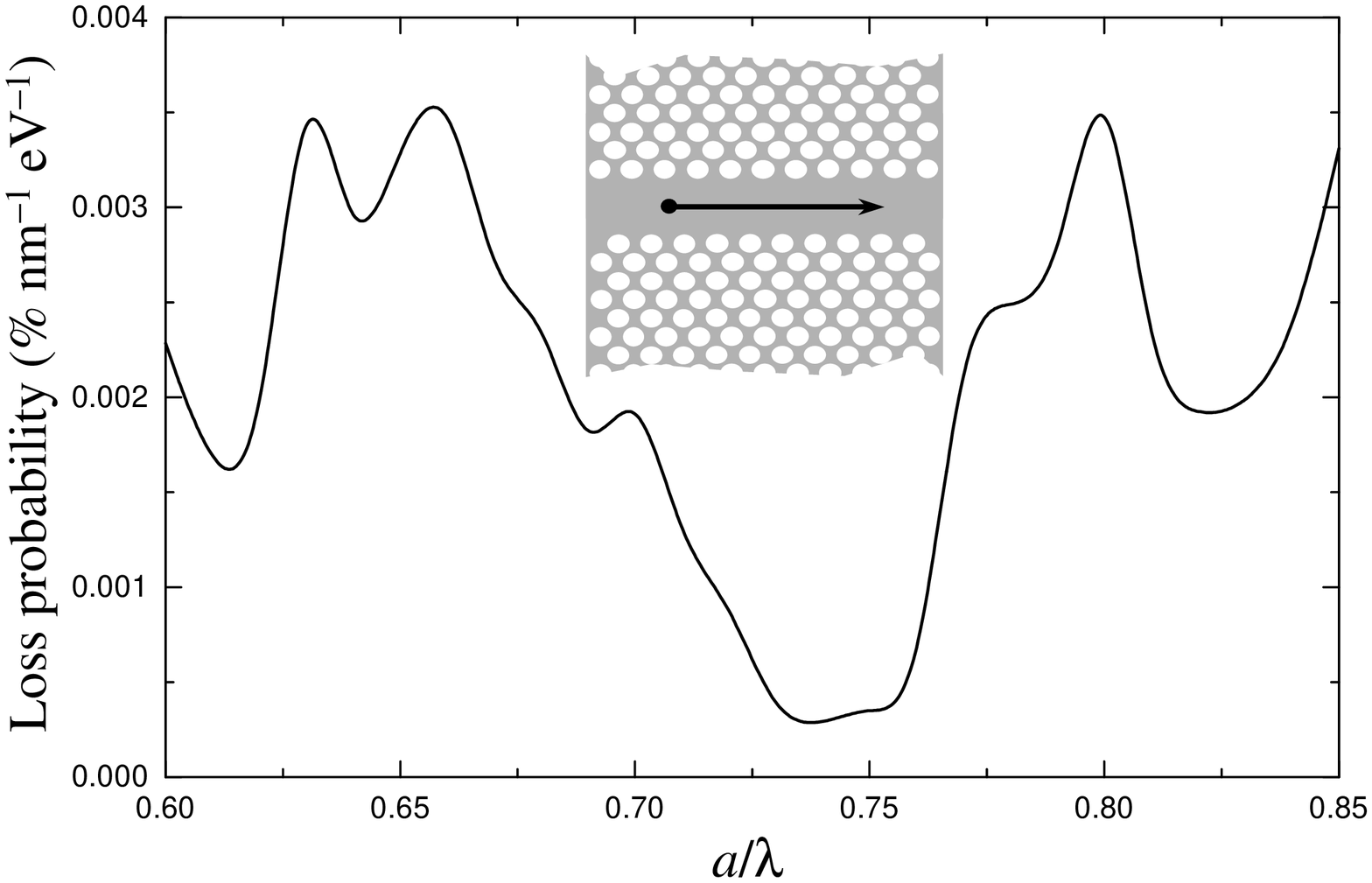}}}
\caption{Energy loss probability for a 100-keV electron traveling
parallel to two photonic crystals with the same parameters as in
Fig.\ \ref{Figf}(c). The spacing between the two crystals is 300
nm, which is filled with Si (this is exaggerated in the inset for
the sake of clarity). The electron is traveling at the same
distance from the two crystals.} \label{Fig7}
\end{figure}

When the lattice constant is comparable to the wavelength, the
idea of an equivalent effective continuous medium is no longer
valid, and one has to rely on the detailed band structure of the
crystal to understand the loss spectra [Figs.\ \ref{Figb},
\ref{Figf}, and \ref{Fig7}].

Finally, the interaction of the electron with the crystal produces
Smith-Purcell radiation, which contributes to the energy loss. If
the crystal is composed of non-absorbing materials, the energy
loss probability and the light emission probability must coincide,
as shown in Fig.\ \ref{Figf} for an inverted Si opal, in which
case the light emitted after transmission through the crystal
shows dips that are directly connected to the presence of photonic
band gaps. For electrons moving in a region surrounded by photonic
crystals a dip is also observed in the loss probability within the
gap energies (Fig.\ \ref{Fig7}).

Our hope is that the present work can provide a stimulus to use
fast electrons in the analysis of photonic crystals as a way to
bring a source of evanescent radiation (the electromagnetic field
of the electron in vacuum) into close contact with the crystal and
also to probe regions of the crystal that would not be easily
accessible to other sources of external electromagnetic radiation.


\acknowledgments

The authors gratefully acknowledge help and support from the
Basque Departamento de Educaci\'{o}n, Universidades e
Investigaci\'{o}n, the University of the Basque Country UPV/EHU
(contract No. 00206.215-13639/2001), and the Spanish Ministerio de
Ciencia y Tecnolog\'{\i}a (contract No. MAT2001-0946).




\begin{references}
\bibitem[*]{email}
E-mail: jga@sw.ehu.es
\end{references}

\begin{references}

\bibitem{M82_1}
L.~D. Marks, Solid\ State\ Commun. {\bf 43},  727  (1982).

\bibitem{C82_1}
J.~M. Cowley, Phys.\ Rev.\ B {\bf 25},  1401  (1982).

\bibitem{HM85_1}
A. Howie and R.~H. Milne, Ultramicroscopy {\bf 18},  427  (1985).

\bibitem{RH88_1}
R.~H. Ritchie and A. Howie, Phil.\ Mag.\ A {\bf 58},  753  (1988).

\bibitem{EP75_1}
P.~M. Echenique and J.~B. Pendry, J.\ Phys.\ C {\bf 8},  2936
(1975).

\bibitem{FE85_1}
T.~L. Ferrell and P.~M. Echenique, Phys.\ Rev.\ Lett. {\bf 55},
1526  (1985).

\bibitem{RZE92_1}
A. Rivacoba, N. Zabala, and P.~M. Echenique, Phys.\ Rev.\ Lett.
{\bf 69},  3362
   (1992).

\bibitem{DF94_1}
B.~T. Draine and P.~J. Flatau, J. Opt. Soc. Am. A {\bf 11},  1491
(1994).

\bibitem{GH98_1}
{F. J. Garc\'{\i}a de Abajo} and A. Howie, Phys.\ Rev.\ Lett. {\bf
80},  5180 (1998); Phys.\ Rev.\ B {\bf 65},  115418 (2002).

\bibitem{G99_2}
{F. J. Garc\'{\i}a de Abajo}, Phys.\ Rev.\ Lett. {\bf 82},  2776
(1999); Phys.\ Rev.\ B {\bf 59},  3095 (1999); Phys.\ Rev.\ E {\bf
61},  5743 (2000).


\bibitem{Y87_1}
E. Yablonovitch, Phys.\ Rev.\ Lett. {\bf 58},  2059  (1987).

\bibitem{JW90_1}
S. John and J. Wang, Phys.\ Rev.\ Lett. {\bf 64},  2418  (1990).

\bibitem{HMT02_1}
S.~D. Hart, G.~R. Maskaly, B. Temelkuran, P.~H. Prideaux, J.~D.
Joannopoulos,
  and Y. Fink, Science {\bf 296},  510  (2002).

\bibitem{LCH98_1}
S.-Y. Lin, E. Chow, V. Hietala, P.~R. Villeneuve, and J.~D.
Joannopoulos,
  Science {\bf 282},  274  (1998).

\bibitem{FMR93_1}
L. Fu, P.~B. Macedo, and L. Resca, Phys.\ Rev.\ B {\bf 47},  13818
(1993).

\bibitem{FR93_1}
L. Fu and L. Resca, Phys.\ Rev.\ B {\bf 47},  16194  (1993).

\bibitem{BF95_1}
R.~G. Barrera and R. Fuchs, Phys.\ Rev.\ B {\bf 52},  3256
(1995).

\bibitem{MBF98_1}
C.~I. Mendoza, R.~G. Barrera, and R. Fuchs, Phys.\ Rev.\ B {\bf
57},  11193
  (1998).

\bibitem{PM94_1}
J.~B. Pendry and {L. Mart\'{\i}n-Moreno}, Phys.\ Rev.\ B {\bf 50},
5062
  (1994).

\bibitem{PM92_1}
J.~B. Pendry and A. MacKinnon, Phys.\ Rev.\ Lett. {\bf 69},  2772
(1992).

\bibitem{SP53_1}
S.~J. Smith and E.~M. Purcell, Phys.\ Rev. {\bf 92},  1069
(1953).

\bibitem{OY01_1}
K. Ohtaka and S. Yamaguti, Optics and Spectroscopy {\bf 91},  506
(2001).

\bibitem{BGxx_1}
L. A. Blanco and {F. J. Garc\'{\i}a de Abajo}, Surf. Sci. (in
press).

\bibitem{evanescentwaves}
Within the retarded response formalism used here, the imaginary
part of $k_h$ is positive. Thus, ${\rm Re}\{\Gamma_Q\}$ and ${\rm
Im}\{\Gamma_Q\}$ must have opposite signs. Now, with our choice of
${\rm Re}\{\Gamma_Q\}>0$, and for $\epsilon_h$ positive and real,
in which case the imaginary part of $k_h$ is infinitesimal (e.g.
in vacuum), one can distinguish too kinds of waves: (i) for
$Q>k_h$, $\Gamma_Q$ is real and positive and then $\Kb_\Qb^+$
represents the momentum of an evanescent plane wave that decays
with increasing $z$; (ii) for $Q<k_h$, $\Gamma_Q$ is imaginary and
$\Kb_\Qb^+$ is the momentum of a propagating plane wave that moves
without attenuation towards the $z>0$ direction.

\bibitem{SYM98_1}
N. Stefanou, V. Yannopapas, and A. Modinos, Comput. Phys. Commun.
{\bf 113}, 49
  (1998); {\bf 132}, 189 (2000).

\bibitem{Gxx_1}
{F. J. Garc\'{\i}a de Abajo} (in preparation).

\bibitem{J75_1}
J.~D. Jackson, {\em Classical Electrodynamics} (Wiley, New York,
1975).

\bibitem{M04_1}
J. C. Maxwell-Garnett, Philos. Trans. R. Soc. London A {\bf 203},
385 (1904);
  {\bf 205}, 237 (1906).

\bibitem{LWA80_1}
W. Lamb, D.~M. Wood, and N.~W. Ashcroft, Phys.\ Rev.\ B {\bf 21},
2248
  (1980).

\bibitem{CKK90_1}
R.~I. Cukier, J. Karkheck, S. Kumar, and S.~Y. Sheu, Phys.\ Rev.\
B {\bf 41},
  1630  (1990).

\bibitem{BD92_1}
D.~J. Bergman and K.-J. Dunn, Phys.\ Rev.\ B {\bf 45},  13262
(1992).

\bibitem{MP95_1}
L. Mart\'{\i}n-Moreno and J.~B. Pendry, Nucl.\ Instrum.\ Methods\
B {\bf 96},
  565  (1995).

\bibitem{MZT00_1}
H. Ma, B. Zhang, W.~Y. Tam, and P. Sheng, Phys.\ Rev.\ B {\bf 61},
962
  (2000).

\bibitem{SSM02_1}
D.~R. Smith, S. Schultz, P. Markos, and C.~M. Soukoulis, Phys.\
Rev.\ B {\bf
  65},  195104  (2002).

\end{references}
\end{document}